\documentclass[aps,twocolumn]{revtex4}
\usepackage{graphicx,epsfig,graphics,amssymb,amsmath}

\def\d{{\rm d}}

\def\fs{f_{\rm slip}}
\def\t{\theta}
\def\deg{$^\circ$}

\def\t{\theta}
\def\be{\begin{equation}}
\def\ee{\end{equation}}
\def\cinf{c_{\infty}}
\def\besub{\begin{subeqnarray}}
\def\eesub{\end{subeqnarray}}

\begin{document}

\title{\textbf{Shear-dependent apparent slip on hydrophobic surfaces: The Mattress Model}}
\author{Eric Lauga \& Michael P. Brenner}
\affiliation{Division of Engineering and Applied Sciences, Harvard
University,\\ 29 Oxford Street, Cambridge, MA 02138.}
\date{\today}

\begin{abstract}
Recent experiments (Zhu \& Granick (2001) {\it Phys. Rev. Lett.}
{\bf 87} 096105) have measured a large shear dependent fluid slip
at partially wetting fluid-solid surfaces. We present a simple
model for such slip, motivated by the recent observations of
nanobubbles on hydrophobic surfaces. The model considers the
dynamic response of bubbles to change in hydrodynamic pressure due
to the oscillation of a solid surface. Both the compression and
diffusion of gas in the bubbles decrease the force on the
oscillating surface by a ``leaking mattress'' effect, thereby
creating an apparent shear-dependent slip. With bubbles similar to
those observed by atomic force microscopy to date, the model is
found to lead to force decreases consistent with the experimental
measurements of Zhu \& Granick.

\end{abstract}
\maketitle

\section{Introduction}

The validity of the no-slip boundary condition is at the center of
our current understanding of fluid mechanics. It remains however
an assumption whose microscopic validity has been widely debated
\cite{Goldstein}. The widespread acceptance of the no-slip
condition is based on a historical record of outstanding agreement
between theories and experiments. It is commonly agreed that the
no-slip condition results from inevitable microscopic roughness,
which causes enough viscous dissipation to effectively bring the
fluid to rest near the surface
\cite{Richardson,Jansons,Granick2002}. Remarkably, this
explanation is independent of the nature of the solid and the
liquid, contrary to ideas first proposed by Girard (see
\cite{Goldstein}).

The development of small devices has recently prompted a
reexamination of fluid slip on length scales of nanometers and
microns, both experimentally
\cite{Granick2002,Pit,Baudry,Craig,Granick,Bonaccurso,Cheng,Meinhart,Cottin}
and theoretically
\cite{Nature,Barrat,BrennerPRE,Koplik,Denniston,Lauga,deGennes}.
The degree of slip is usually quantified by a slip length $
\lambda={U_s}/{\dot{\gamma}}$, where $U_s$ is the slip velocity
and $\dot{\gamma}$ is the liquid strain rate evaluated at the
surface; equivalently, $\lambda$ is the distance below the solid
surface where the velocity extrapolates linearly to zero
\cite{Navier}. In experiments, slip is usually found when the
liquid partially wets the solid surface; measured slip lengths
span four orders of magnitude, from molecular sizes to microns,
and are usually shear-dependent in squeeze flow experiments, with
$\lambda$ an increasing function of $\dot{\gamma}$. In particular,
Zhu \& Granick \cite{Granick} reported squeeze flow experiments,
in which two crossed cylinders oscillate about a fixed average
distance. By measuring the viscous resistance, Zhu \& Granick
extracted the slip length over a wide range of oscillation
amplitudes. These experiments lead to the largest shear-dependent
slip lengths yet (up to $\sim$ 2 $\mu$m).

The origin of this large shear dependent slip is heretofore
mysterious. Nanobubbles have recently been observed on hydrophobic
surfaces, using atomic force microscopy, with typical thickness
$h\sim 10$ nm, typical radius $R\sim 50-100$ nm and high surface
coverage \cite{nano1,nano2,nano3,nano4}.  Although the origin of
these bubbles is unclear and skepticism remains in the community
about their existence, they have been often invoked as a possible
origin of the so-called hydrophobic attraction
\cite{hydro0,hydro2,hydro3,hydro5,hydro6} and their existence
points to a possible picture for such large slip
\cite{nano2,deGennes}.

It is well known that there is in general a non-zero velocity at a
liquid-gas interface, and therefore it is natural to wonder
whether the existence of such a gas layer at the solid surface is
sufficient to explain the experiments. When a fluid of viscosity
$\eta_1$ adjoins a  layer of fluid of thickness $h$ with smaller
viscosity $\eta_2$, the discontinuous strain rate at the
fluid-fluid interface results in an apparent slip with slip length
\begin{equation}\label{length}
\lambda=h\left(\frac{\eta_1}{\eta_2} -1\right) \cdot
\end{equation}
Choosing $\eta_1/\eta_2=50$ appropriate for a gas-water interface
leads to slip lengths as large as $500$ nm. This estimate is
however independent of the interfacial shear and therefore unable
to explain the squeeze flow experiments; it also overestimates the
slip length in the case of bubbles, as is discussed in section
\ref{comparison}. Note however that similar arguments are
consistent with data from pressure-driven flow experiments where
reported slip lengths to date are essentially shear-independent
\cite{Cheng,Meinhart,Schnell,Churaev,Watanabe,Lauga}.

In this article, we will assume  bubbles exist on hydrophobic
surfaces and will calculate their {\sl dynamic} response to an
imposed oscillatory shear. In an oscillatory squeeze flow
experiment \cite{Granick}, we argue that the pressure fluctuations
in the fluid cause the bubbles to act as a ``leaking mattress'',
with both compression and dilation of the gas in the bubble, as
well as diffusion of gas into (and out of) the bubble. As the
solid sphere oscillates, this periodic in-phase response of the
bubbles sizes reduces the amount of liquid necessary to be
squeezed out of the gap and thereby the force on the moving
sphere, creating an apparent slip. Our calculations indicate that
the magnitude of this apparent slip is consistent with the
observations of Zhu \& Granick. We present the details of our
model in the next section and discuss the comparison with the
experiment in section \ref{comparison}.

\section{Influence of bubbles on force measurements}
\label{calculation}

A typical oscillatory squeeze flow experiment is shown in
Figure~\ref{figure}. A sphere of radius $a$ oscillates with
velocity $V_S$ in a viscous liquid of viscosity $\eta$ at a
distance $D$ of a planar surface (equivalently, the surfaces can
be two crossed cylinders). The two surfaces are assumed to have
the same physico-chemical properties. If no bubbles are present
and the no slip boundary condition is satisfied on both surfaces,
the lubrication force opposing the motion of the sphere is given
by the Reynolds equation
\begin{equation}
 F(t)={\bf e}_z\cdot {\bf F}(t)=- \frac{6\pi\eta a^2}{D} V_S\triangleq F_{\rm lub}. \label{viscous}
\end{equation}
If however flow occurs on the surfaces with a slip length
$\lambda$, the viscous force is decreased by an amount $\fs$
\cite{Vino95} given by
\begin{equation}\label{f1}
\fs=\frac{D}{3\lambda} \left[\left(1+\frac{D}{6\lambda} \right)
\ln\left(1+\frac{6\lambda}{D}\right)-1\right]\cdot
\end{equation}
Equation \eqref{f1} is used experimentally to infer effective slip
lengths: the experimental  viscous force $F_{\rm exp}$  is
compared to the theoretical no-slip result $F_{\rm lub}$ and any
difference is interpreted as fluid slip, with a slip length
$\lambda$ corresponding to $ \fs=F_{\rm exp}/F_{\rm lub} $.

Let us now {\sl assume} that the solid surface is covered with a
percentage $\phi$ of identical gas bubbles (Figure \ref{figure}),
and determine how the bubbles modify the dynamic response.
Although this assumption has been made by previous authors
\cite{hydro2,hydro3,hydro5,hydro6,nano2,deGennes},  the physical
mechanism responsible for such bubbles is unknown. Simple
estimates indicate that small bubbles are short lived in solution
\cite{lifetime}. However, stable bubbles could arise from any
number of possibilities that are known to prolong bubble
lifetimes, including surfactants, surface heterogeneities, or
local supersaturation of dissolved gas \cite{attard}. In this
paper, we are interested in understanding whether the dynamic
response of hypothesized bubbles is sufficient to rationalize slip
experiments.

\subsection{Total force}

The total force $F(t)$ resisting the oscillatory motion of the
area $S$ of the sphere has two components: (1) a viscous
lubrication force $F_h$, due to hydrodynamic pressure fluctuations
and acting on an area $(1-\phi)S$ and (2) an elastic bubble force
$F_b$, due to pressure fluctuations inside the bubbles and acting
on an area $\phi S$. The total force is therefore given by
\begin{equation}\label{}
F(t) =(1-\phi)F_h+\phi F_b,
\end{equation}
where
\begin{equation}\label{forces}
F_h=(p-p_0)S,\quad F_b=(p_b-p_{eq})S.
\end{equation}
Here $p$ and $p_b$ ($p_0$ and $p_{eq}$) denote the (equilibrium)
pressures in the liquid and the bubbles respectively. Moreover,
since $D\ll a$, the surface $S$ is given by $S\approx\pi aD$.

\subsection{Lubrication force}

Let us first calculate the hydrodynamic force $F_h$.

The presence of bubbles modifies equation \eqref{viscous} in two
ways. First, flow occurs over a distribution of bubbles located on
an otherwise no-slip surface, so the viscous force is reduced by
an amount $\fs$ given by equation \eqref{f1}, where $\lambda$ is
the appropriate effective slip length for flow over a distribution
of bubbles \cite{Philip,Lauga}.

Second, the size of the bubbles changes in time in response to
pressure fluctuations in the liquid. This volume effect will
modify the amount of liquid necessary to be squeezed out of the
gap at each cycle of the oscillations, hence the viscous force.
Consequently, bubble dynamics has to be subtracted from the
forcing $V_S$ and the hydrodynamic lubrication $F_h$ force is now
given by a modified Reynolds equation
\begin{equation}
F_h=-\fs\frac{6\pi\eta a^2}{D} \left(V_S-2\frac{{\rm d} h}{{\rm
d}t}\right) \label{viscousforce}
\end{equation}
where $h$ is an average bubble thickness on each surface and the
factor 2 accounts for the fact that each surface is covered with
bubbles.

\begin{figure}[t]
\centering
\includegraphics[width=.49\textwidth]{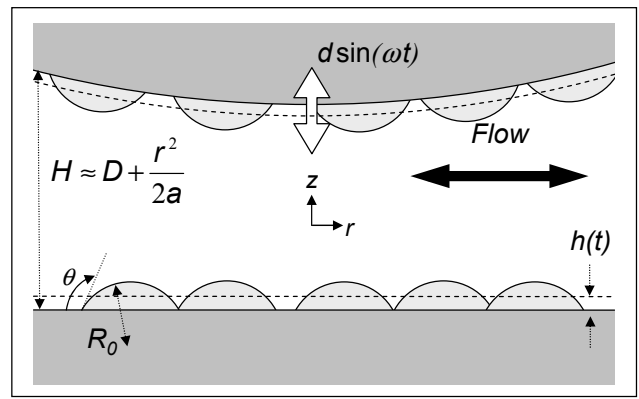}
\caption{Typical squeeze flow experiment: a solid sphere of radius
$a$ is oscillated in a liquid at a distance $D\ll a$ of a smooth
solid surface with amplitude $d\ll D$ and frequency $\omega$. The
surfaces are covered by microscopic gas bubbles of contact angle
$\t$ and radius of curvature $R_0$. The set of bubbles is
approximated by a gas layer of time evolving thickness $h(t)$.}
\label{figure}
\end{figure}

\subsection{Rate of change of bubble height}

In order to calculate $\d h / d t$ in equation
\eqref{viscousforce}, let us now consider the dynamics of the
bubbles. We assume the bubbles are undeformed by viscous stresses
and remain spherical, with radius of curvature $R(t)$ and interior
angle ($\pi-\theta$) (see Figure~\ref{figure}). We  neglect
interactions between bubbles. We expect  $h(t)$ to depend
explicitly on the forcing on the bubbles, {\it i.e.} $F_b$.

At the small frequencies typical of squeeze flow experiments
(1-100 Hz), the gas is isothermal, so the pressure in the bubble
changes via the ideal gas law
\begin{equation}\label{ideal}
\frac{p_b(t)V(t)}{m(t)} =\frac{p_{eq}V_0}{m_0},
\end{equation}
where $V$ and $m$ denote the volume and the mass of a single
bubble. The average thickness of the gas layer is defined as
$h(t)=nV(t)$, where $n$ is the number density of bubbles on the
solid surface, so that equation \eqref{ideal} can be rewritten as
\begin{equation}\label{new ideal}
\frac{p_b(t) h(t)}{m(t)}=\frac{p_{eq}h_0}{m_0}\cdot
\end{equation}
Combining the time-derivative of \eqref{new ideal} with
$F_b=(p_b-p_{eq})S$ and linearizing  around $\{p_b,h,m\}\sim
\{p_{eq},h_0,m_0\}$, we obtain the equation for the rate
  of change of the mean bubble height $h$
\begin{equation}\label{changeh}
\frac{\d h}{\d t}=\frac{h_0}{m_0}\frac{\d m}{\d
t}-\frac{h_0}{p_{eq}S}\frac{\d F_b}{\d t}
\end{equation}
We thus have that the rate of change of $h$ is the sum of a rate
of change  governed by gas diffusion plus a second contribution
due to the gas compressibility.

We now consider the rate of gas diffusion from the bubble.  In our
model for the oscillatory squeeze flow experiments
\cite{Baudry,Granick,Granick2002}, bubbles lose mass by both
vertical diffusion across the liquid layer and radial diffusion
along the apparatus; because of the scale separation $D\ll a$,
these two processes require separate treatment.

Let us first consider the case of vertical diffusion. Since for
most common gases, $\kappa\sim 10^{-9}$~m$^2$/s,  the vertical
Peclet number $Pe_v = D^2 \omega/\kappa$ is much smaller than
unity: on the experimental time scale $\omega^{-1}$, the bubbles
are approximately in instantaneous vertical diffusive equilibrium.
The dissolved gas concentration above the bubble is therefore
uniform throughout the liquid gap and is given, in the case of
small amplitude oscillations, by Henry's law $c=p_b\cinf/p_{eq}$.
The mass $\tilde{m}$ of gas necessary to fill the liquid gap at
this concentration is equal to the gap thickness $(D-2h)$ times
the change in concentration $\cinf(p_b/p_{eq}-1)$ times the area
in the liquid which is influenced by the bubble, {\it i.e.} $1/n$
. Linearizing around $h\approx h_0$ and combining with
$p_{eq}/\cinf=p_0/c_0$, we get that $\tilde{m}$ is proportional to
the bubble force $F_b$
\begin{equation}\label{}
\tilde{m}=\frac{c_0 (D-2h_0)}{n p_{0}S}F_b\cdot
\end{equation}
Consequently, the total rate of change of $m$ is given by
\begin{equation}\label{vertical}
\frac{\d m}{\d t}=\frac{\d m_r}{\d t}-\frac{c_0 (D-2h_0)}{n
p_0S}\frac{\d F_b}{\d t}
\end{equation}
where $\d m_r /\d t$ is the rate of change in the bubble mass
governed by gas diffusion in the (slowly varying) radial direction
of the apparatus; let us now evaluate this contribution.

In contrast to  the vertical case, the radial oscillatory Peclet
number, $Pe_r= L^2\omega / \kappa=aD\omega / \kappa$, is of order
unity or larger, so that radial diffusion has to be accounted for
explicitly. Assuming the dissolved gas is in vertical diffusive
equilibrium, the time rate of change of the mass of a gas bubble
${\d {m_r}}/{\d t}$ is given by a flux integral on the bubble
surface $S_b$
\begin{equation}\label{}
\frac{\d {m_r}}{\d t} = \kappa\int_{S_b} {\bf n}\cdot \nabla c
\,{\rm d}S = \kappa R^2 I(\t)\frac{\partial c}{\partial r},
\label{masschange}
\end{equation}
where the assumption od spherical cap bubble  implies that
$I(\t)={\pi}(2(\pi-\t)+\sin 2\t)/2$.

In general, the radial concentration of dissolved gas $c(r,t)$
verifies an advection-diffusion equation with shear dependent
diffusivity \cite{TaylorDispersion}. However, for the small
amplitude oscillation in \cite{Granick}, both advection and Taylor
dispersion are negligible, and $c(r,t)$ satisfies a pure diffusion
equation. We finally approximate radial concentration gradient by
a simple linear law $ \partial c / \partial r \approx (\cinf-c)/
L_r$ where $L_r\approx (\kappa/\omega)^{1/2}\approx L/Pe_r^{1/2}$
is the typical (shear dependent) radial gradient length scale.
Equation~(\ref{masschange}) together with Henry's law leads
therefore to a linear relation between the rate of change
$\dot{m_r}$ and the bubble force
\begin{equation}
\frac{\d m_r}{\d t}=- \frac{\kappa R_0^2I(\t)c_0 }{p_{0}SL_r} F_b
\,\cdot \label{c1}
\end{equation}

Combining $m_0=\rho_0 h_0/n$
with~(\ref{changeh}),~(\ref{vertical}) and~(\ref{c1}), we finally
obtain that the mean bubble height $h$ satisfies the differential
equation
\begin{equation}
\frac{\d h}{\d t}= -k_1 F_b-k_2 \frac{\d F_b }{\d t} \label{ODEh},
\end{equation}
where $(k_1,k_2)$ are given by
\begin{eqnarray}
k_1&=&\frac{n\kappa R_0^2I(\t)c_0}{\rho_0p_0SL_r} ,\label{c}\\
k_2&=& \frac{c_0 h_0}{\cinf p_0 S}\left(1+\frac{ \cinf(D-2h_0)
}{\rho_0h_0}\right)\cdot\label{k2}
\end{eqnarray}

\subsection{Bubble force}

We finally need to calculate the bubble force $F_b$ in order to
close the system of equations \eqref{viscousforce} and
\eqref{ODEh}. This in general requires understanding the
(equilibrium or nonequilibrium) mechanism responsible for the
presence of these long-lived bubbles. However, we can bypass this
unknown physics by assuming without loss of generality that the
pressure fluctuations in the bubbles and that in the liquid are
proportional
\begin{equation}\label{}
\Delta p_b =\alpha\, \Delta p,
\end{equation}
where $\alpha$ is an unknown constant. Hence, bubble and
hydrodynamic forces are proportional $F_b =\alpha F_h$, and the
total force on the sphere can be expressed as
\begin{equation}\label{twoforces}
F=\left(1-\phi+\alpha\phi \right)F_h
=\left(\frac{1-\phi+\alpha\phi }{\alpha} \right)F_b.
\end{equation}

\subsection{Total force on the sphere}

We can now combine \eqref{viscousforce}, \eqref{ODEh} and
\eqref{twoforces} to express the total force opposing the motion
of the sphere $F$. We obtain
\begin{equation}\label{first}
F(t)=-\frac{\delta}{2}(1-\phi+\alpha\phi)
\left(V_S+2k_1F_b+2k_2\frac{\d F_b}{\d t}\right),
\end{equation}
where$\delta={12\pi\eta a^2\fs}/{D}$. Since $F$ and $F_b$ are
related by equation \eqref{twoforces}, \eqref{first} can be
transformed into an ordinary differential equation for $F$,
\begin{equation}
 F(t)=-\frac{\delta}{2} \left(\widetilde{V_S}+
 2\alpha k_1 F+2\alpha k_2
\frac{\d F}{\d t}\right),\label{ODEF}
\end{equation}
where $\widetilde{V_S} = (1-\phi+\alpha\phi)V_S$. For an
oscillating sphere velocity $V_S={\d } \left( d \sin\omega t\right
)/\d t$, the periodic solution to \eqref{ODEF} is given by
\footnote{The general solution to \eqref{ODEF} also an
exponentially decaying transient occurring on a time scale $
\tau\sim \delta k_2/(1+\delta k_1)$ with $\tau  \sim 1$ ms or less
in experiments such as \cite{Granick}.}
\begin{eqnarray}\label{sol}
\frac{F(t)}{F_{\rm lub}}&=&\fs\frac{(1-\phi+\alpha\phi)(1+\delta
\alpha
k_1)}{(1+\delta \alpha k_1)^2 +(\delta\omega \alpha k_2)^2}\\
\nonumber & & \times \left(\cos \omega t +\frac{\delta \omega
\alpha k_2}{1+\delta\alpha k_1}\sin\omega t\right)\cdot
\end{eqnarray}

\subsection{What is the value of $\alpha$?}

By comparing our model \eqref{sol} to the results of squeeze flow
experiments, we find that the only choice consistent with
available data at large separation distances is $\alpha\approx 1$.

To see this, consider equation \eqref{sol} in the limit of large
separations between the sphere and the planar surface $D$. Since
equation \eqref{f1} shows that $\fs \sim 1$ when $D$ is large, we
get $\delta\sim D^{-1}$. Moreover, $S\sim D$ so that, from
\eqref{c} and \eqref{k2}, we obtain $\delta k_1 \sim D^{-2}$ and
$\delta k_2\sim D^{-1}$. Consequently, in the limit of large
separations, we obtain that the ratio of the measured force,
out-of-phase with the sphere displacement, to the expected no-slip
force \eqref{viscous}  is given by
\begin{equation}\label{}
\lim_{D\to \infty}\left(\frac{F}{F_{\rm lub}}\right)=
1-\phi+\alpha\phi.
\end{equation}
Within experimental errors, this ratio is always measured to be
unity \cite{Granick2002,Craig,Granick,Bonaccurso,Cottin}, {\it
i.e.} the expected lubrication no-slip force is recovered for
large separation distances. We therefore need $ 1-\phi+\alpha\phi
\approx 1$ or $\alpha \approx 1$.

We emphasize that this conclusion is reached because we {\it
assume} that the model presented in \ref{calculation} is the major
physical mechanism responsible for the force decrease observed in
experiments such as \cite{Granick}.

\subsection{Final formula for the force ratio}

We obtain from \eqref{sol} that the ratio $f^*$ of the peak force
out-of-phase with the sphere displacement to that expected with
no-slip and no bubbles \eqref{viscous} is given by
\begin{equation}\label{final}
\frac{f^*(\omega)}{\fs}= \frac{1}{1+\left(\delta k_1
+\displaystyle \frac{( \omega \delta k_2)^2}{1+\delta
k_1}\right)}\cdot
\end{equation}
The ``leaking mattress'' model therefore leads to an apparent slip
effect, of dynamic origin. The effect is shear-dependent through
the frequency dependence in \eqref{final}: higher frequency and
therefore higher shear shear rates lead to a larger apparent slip,
in agreement with \cite{Granick2002,Craig,Granick}. The model was
derived under the assumption of small amplitude oscillations $d$,
which consequently does not appear in the final formula for $f^*$.

Note that in the limit of high frequencies  the force ratio
\eqref{final} becomes $ f^*(\omega\to\infty) \sim \omega^{-2}$,
whereas at low frequencies $f^*(\omega\to 0) \approx
\frac{1}{1+\delta k_1}.$ In the low frequency limit $f^*$ is
independent of frequency and depends only on separation distance.
Furthermore, equation \eqref{final} implies that the apparent slip
effect increases with the fluid viscosity, in agreement with
experiments \cite{Craig,Cheng}. It also increases with the size of
the sphere $a$, which might account for the large slip lengths
reported in \cite{Granick} (cm-size spheres) as opposed to other
squeeze flow experiments (usually $\mu$m-size spheres). The model
predicts that the measured overall apparent slip length is
therefore not only a solid/liquid property but depends on the
system size \cite{Lauga}.

We finally note from \eqref{sol} that the total pressure force on
the sphere $F(t)$ also contains a non-zero component in-phase with
the displacement of the sphere, and therefore out-of-phase with
its velocity. If we denote by $g^*$ the ratio of this in-phase
response to the expected no-slip no-bubbles out-of-phase response,
we obtain
\begin{equation}
 g^*=\frac{\delta \omega k_2}{(1+\delta k_1)^2 +(\delta \omega
k_2)^2}\fs=\frac{\delta \omega k_2}{1+\delta k_1}{f^*}
\cdot\label{elastic}
\end{equation}
Equation \eqref{elastic} is a prediction of the effective
elasticity provided by the bubbles to the surface, which would
occur in addition to other in-phase contribution such as
intermolecular forces, and is experimentally testable. The values
of the in-phase responses of the forces were unfortunately not
reported by Zhu \& Granick \cite{Granick}.

\section{Comparison with experiments}
\label{comparison}

We present in this section a quantitative comparison of our model
with the experimental results of Zhu \& Granick in the case of
deionized water, namely the four sets of data presented in Figure
2 of ref \cite{Granick}. The macroscopic water/solid contact angle
in this case was $110$\deg\, the sphere radius was $a=2$ cm and we
assumed that the liquid was saturated with O$_2$ at 25\deg C and 1
atm ($\cinf=c_0$), for which $\rho_0=1.28$ kg/m$^3$,
$c_0=8.3\times 10^{-3}$ kg/m$^3$ and $\kappa=2\times 10^{-9}$
m$^2$/s. As a matter of comparison, we have also summarized in
Table \ref{table} the experimental results of
\cite{nano1,nano2,nano3,nano4} on the typical size, distribution
and morphology of bubbles observed by atomic force microscopy.

\begin{figure}[t]
\centering
\includegraphics[width=.48\textwidth]{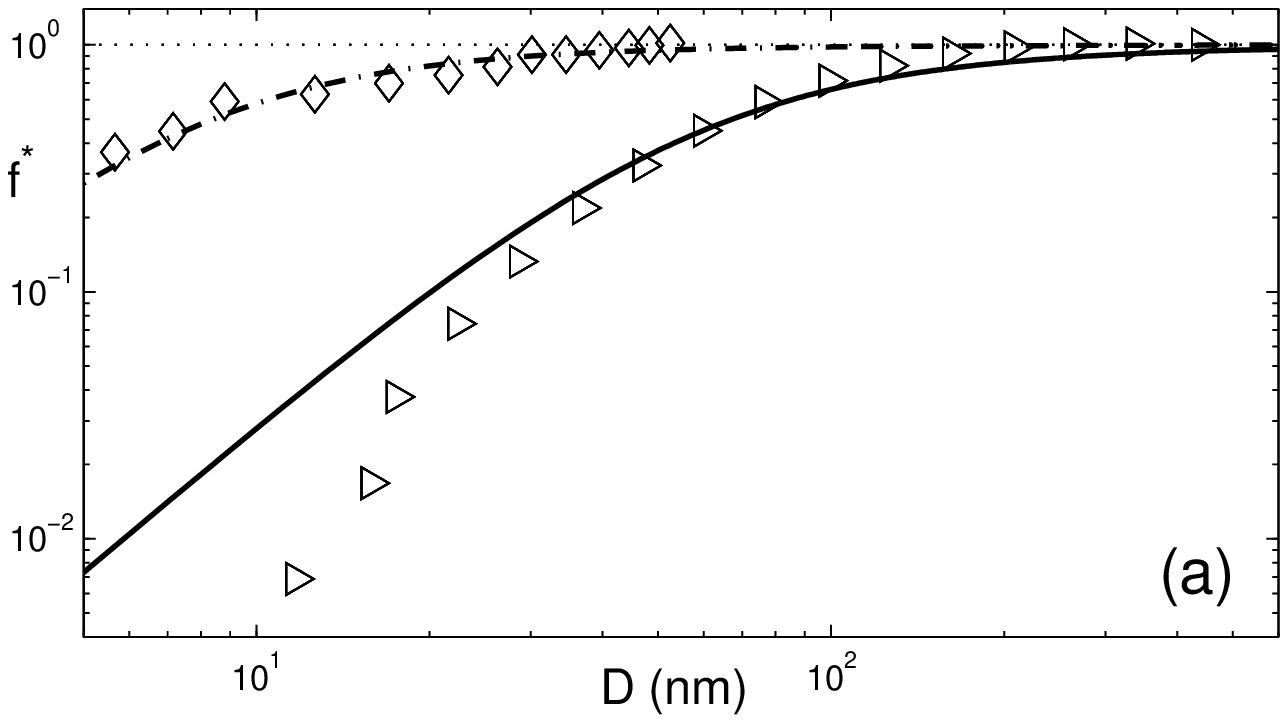}
\includegraphics[width=.48\textwidth]{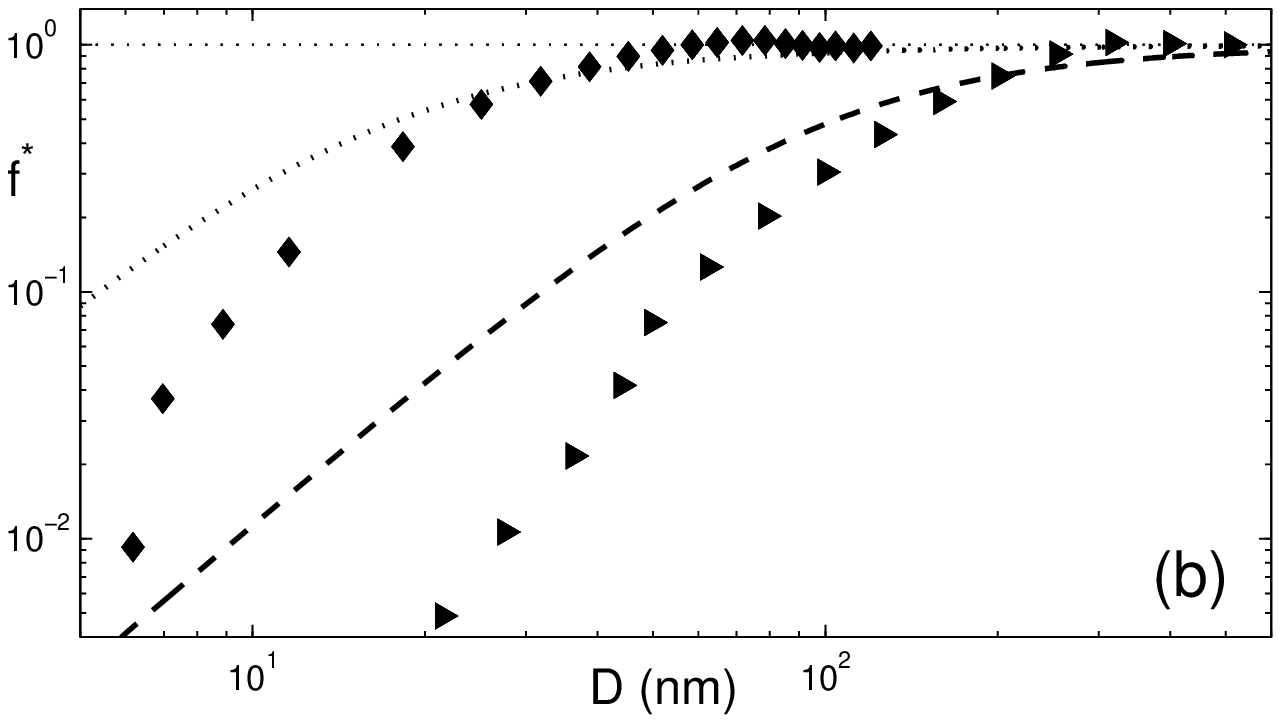}
\caption{Comparison between the experimental data of Zhu \&
Granick (2001) and the dynamic model \eqref{final} with $R_0=10$
nm and $\phi=99\%$. (a): Small amplitude experimental data;
($\lozenge$): measurements for $d=0.5$ nm, $\omega=1$ Hz;
dashed-dotted line: model for $\theta=177$\deg;
($\vartriangleright$): measurements for $d=1.6$ nm, $\omega=10$
Hz; solid line: model for $\theta=132$\deg. (b): Large amplitude
experimental data. ($\blacklozenge$): measurements for $d=6$ nm,
$\omega=1$ Hz; dotted line: model for $\theta=168$\deg;
($\blacktriangleright$): measurements for $d=6$ nm, $\omega=10$
Hz; dashed line: model for $\theta=90$\deg. } \label{exp}
\end{figure}

The ``leaking mattress" model we have presented in the previous
sections has three free parameters which we  fit to the
experimental data: (a) the area fraction of the bubbles on the
surface, $0\leq \phi \leq 1$, (b) the size of the spherical
bubbles, described by their radius of curvature $R_0$ and  (c) the
microscopic contact angle $\theta$ at the bubble level, which
significantly differs from the macroscopic contact angle because
of both intermolecular forces at the nanometer scale. Note that
$\phi$ is related to the area fraction $n$ by the formula
$\phi$=$n\pi R_0^2\sin^2\theta$.

Furthermore, in order to present a meaningful fit to available
data, we require that in each experiment the two layers of bubbles
fit in the gap between the sphere and the plane for  all
separation distance. This is a geometrical constraint written as
$2 h_0 = 2 R_0 (1+\cos \theta) \leq \,\min(D)$ \footnote{The model
also requires specification of the effective slip length $\lambda$
in equation \eqref{f1}.  Since a set of gas bubbles on a surface
resists fluid motion more strongly than a gas layer,  equation
\eqref{length} is not appropriate. We have experimented with
several models (see e.g. \cite{Lauga,Philip}), all of which give
slip length in the range ($\lambda \sim 10-100$ nm depending on
the experiment). Owing to the weak dependence of $f^*$ on the slip
length, the precise value of $\lambda$ has virtually no impact on
the quantitative results of our fits.}

The model \eqref{final} can be well fit to the experiments
\cite{Granick} with appropriate parameter choices. The best fits
are obtained when we choose $R_0 \approx 10$ nm. This is
illustrated in Figure \ref{exp} where the fits are compared the
force ratio from the model to the small (a) and large (b)
amplitude data from Zhu \& Granick \cite{Granick}; the values of
the angles $\theta$ were chosen for each curve to be the best in a
least-square sense and $\phi=99\%$. As expected from the linearity
of our model, the fit to the low amplitude data of \cite{Granick}
is better than that obtained for oscillations of larger amplitude.

\begin{figure}[t]
\centering
\includegraphics[width=.48\textwidth]{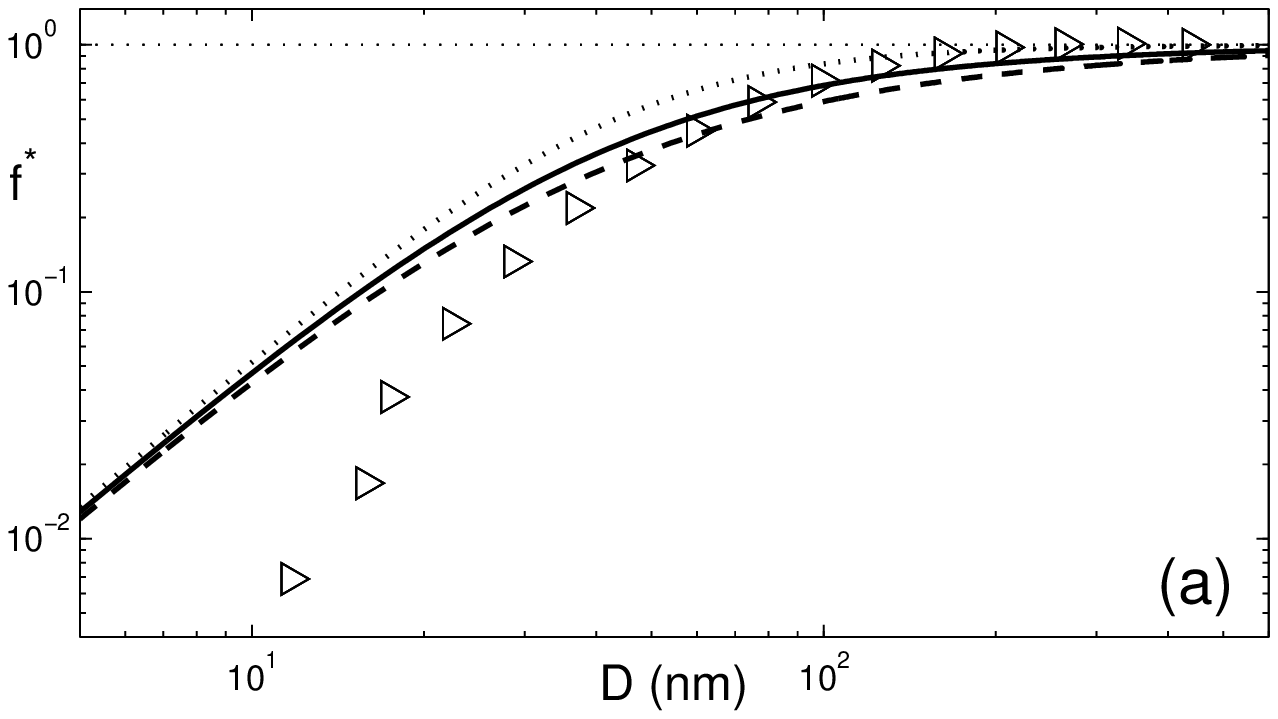}
\includegraphics[width=.48\textwidth]{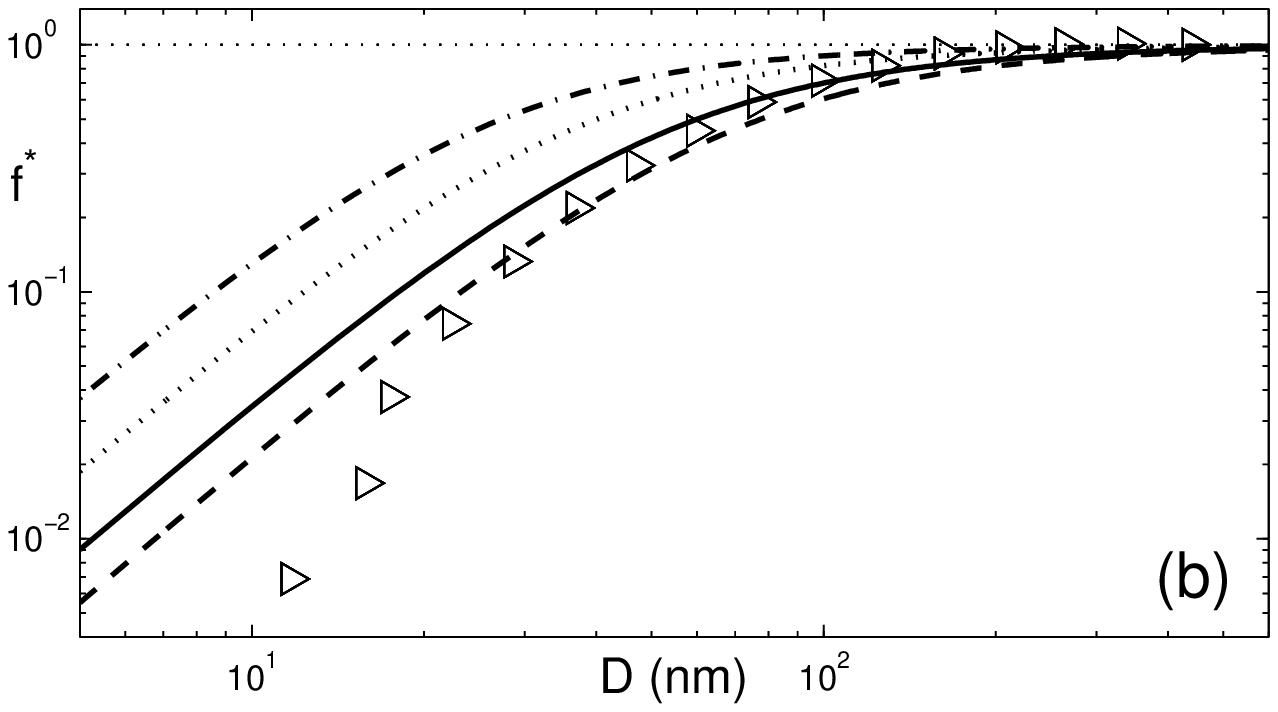}
\includegraphics[width=.48\textwidth]{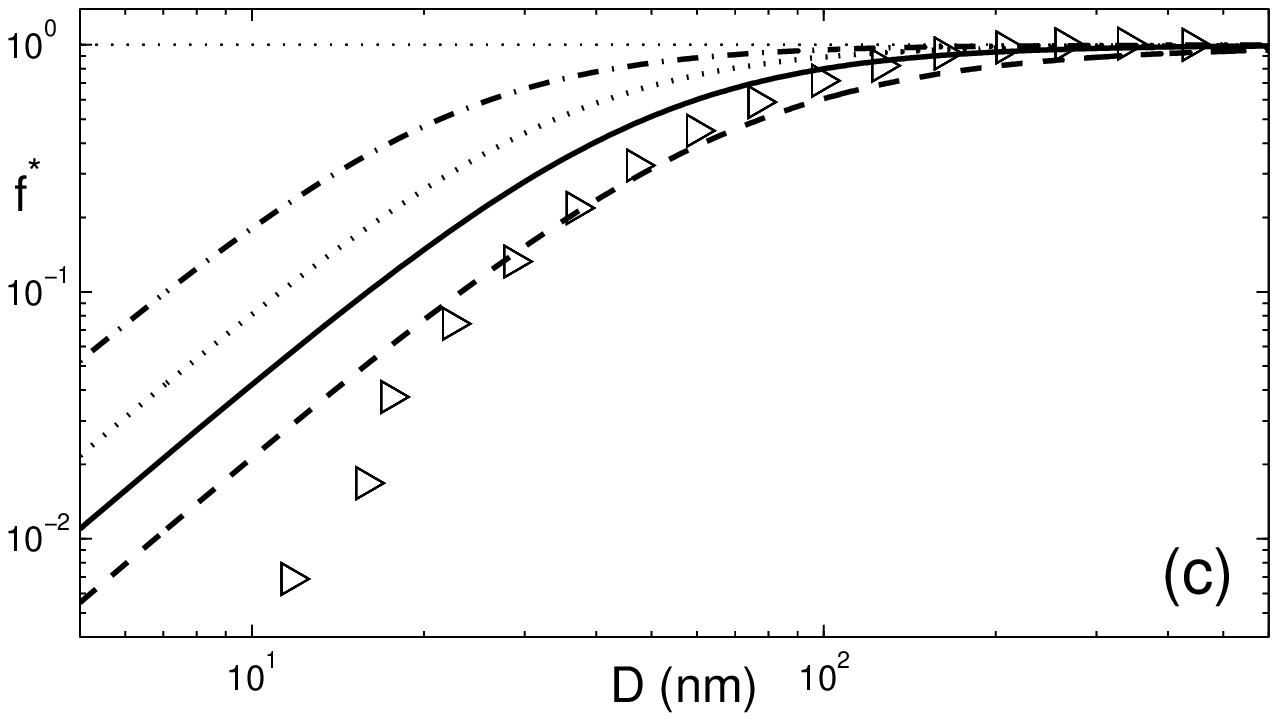}
\caption{Comparison between the experiment from \cite{Granick}
with $d=1.6$ nm and $\omega=10$ Hz ($\vartriangleright$) and the
model for different bubble sizes, contact angles and surface
coverage. (a): Influence of bubble size; model with surface
coverage $\phi=0.99$, contact angle $\theta=150$\deg\,and bubble
sizes $R_0=$1 nm (dotted line), 25 nm (solid line) and 50 nm
(dashed line). (b): Influence of contact angle; model with surface
coverage $\phi=0.99$, bubble size $R_0=10$ nm and contact angles
$\theta=120$\deg\,(dashed line), 140\deg\,(solid line),
160\deg\,(dotted line) and 170\deg\,(dashed-dotted line). (c):
Influence of surface coverage; model with bubble size $R_0=10$,
contact angle $\theta=120$\deg\,and surface coverage $\phi=0.1$
(dashed-dotted line), $\phi=0.25$ (dotted line), $\phi=0.5$ (solid
line) and $\phi=0.99$ (dashed line).}\label{change}
\end{figure}

We explore the influence of the three parameters of our model
($\phi,R_0,\theta$) in Figure \ref{change}a-c for the measurements
from \cite{Granick} with  $d=1.6$ nm and $\omega=10$ Hz.

We first find that the results of our model depend weakly on the
bubble sizes: the results on Figure \ref{change}a are consistent
with the experimental data for a large range of bubble sizes,
$R_0\sim 1$ to 50 nm. These sizes are agreement with the
experimental evidence of bubbles in \cite{nano1,nano2,nano3,nano4}
as summarized in Table \ref{table}, although somewhat smaller. As
a matter of comparison, the data in \cite{nano3} show large
standard deviation (up to 70\%) for the bubble area.

As a difference, we find that the results of our model depend on
both the assumed microscopic contact angle $\theta$ and coverage
of the surface by the bubbles $\phi$. We observe variations in the
contact angles leading to best fit to the four experiments (Figure
\ref{exp}) and also note that we obtain a departure from the best
fit when the angle is chosen to be significantly different (Figure
\ref{change}b). In three experiments out of four, we find that the
microscopic contact is larger than the macroscopic contact angle
$110^\circ$ characterizing the wetting of deionized water on the
surfaces used in \cite{Granick}. This result is consistent with
the data in Table \ref{table} where, in all cases, bubbles were
found experimentally to be flat with microscopic contact angles
larger than the macroscopic wetting angles. The fourth set of data
from \cite{Granick} is found to be consistent with a microscopic
angle of about $90^\circ$. Although this is different from the
data in \cite{nano1,nano2,nano3,nano4}, it is consistent with
theoretical studies which show that intermolecular forces lead to
microscopic contact angles which are always closer to 90\deg~than
their macroscopic counterpart \cite{hocking}. Furthermore, we note
that electrical effects are known to have significant impact on
contact angles of bubbles and drops \cite{electrowetting,chou}.

Finally, we find that our model is  consistent with the
experimental data when the surface coverage is assumed to be large
and almost equal to 100\% (see Figure \ref{change}c). This result
compare well with the available data on bubbles where, in three
out of four studies \cite{nano2,nano3,nano4}, the bubbles were
found to cover almost entirely the solid surface. As a difference,
the pictures in \cite{nano1} show bubbles with lower surface
coverage. We also note that our previous study of slip in
pressure-driven flow experiments lead to a similar conclusion: in
order for surface-attached bubbles to be responsible for the
measured effective slip, surface coverage of almost 100\% was
necessary \cite{Lauga}.

\section{Conclusion}

We have explored in this paper the consequences of the presence of
nanobubbles on the surfaces where squeeze flow experiments are
performed. We have shown that, within the framework of a simple
stabilizing model, the time-dynamics of bubbles always leads
naturally to a shear-dependent decrease in the measured viscous
force by a ``leaking mattress'' effect. The effect was found to
increase with viscosity of the fluid and the size of the sphere,
in agreement with earlier experimental results.

We emphasize that this mechanism is of {\it dynamic} origin, and
is not a consequence of the microscopic slip at the bubble
surfaces; in particular, we argue that this is why shear-dependent
slip length have not been reported by investigations of slip in
pressure-driven flow experiments to date, where no oscillatory
pressure is present to trigger an effect similar to the one
proposed here. Also, the mechanism we propose should also apply to
squeeze flow experiments performed on super-hydrophobic surfaces
such those reported in \cite{Onda} with small air bubbles trapped
on fractal surfaces (see also \cite{Bico}).

Assuming the presence of bubbles, the calculations on the model
have been performed with several simplifying assumptions and, in
particular, additional contributions to the sets of coefficients
$(k_1,k_2)$ could come from bridging bubbles, large amplitude
oscillations of the solid sphere or bubble interactions,
deformation or displacement on the solid surface.

We have then presented a comparison between the results of our
model when applied to the experiments of Zhu \& Granick. We found
that our model gives results which are in agreement with the force
decrease measured experimentally, for bubble features which are
consistent with available experimental data on nanobubbles (bubble
size $R_0\sim 10 $nm, large microscopic contact angles, large
surface coverage). Finally, a formula has been proposed for the
(additional) effective elasticity provided by the bubbles to the
solid surface.

We note that our study does not rule out the possibility of
bubbles with dynamically selected sizes. It has been reported
experimentally in \cite{Yakobov} that the jump-in distance between
two hydrophobic surfaces in water, believed to be due to the
presence of bubbles, depended on the history of the sample;
performing the experiment several times lead to changes in the
jump-in distances over time which was found to remain constant
only after a few periods. A similar scenario could be envisioned
in experiments such as \cite{Granick}.

To conclude, we present a simple prediction based on the results
of our model. If a squeeze flow experiment was performed with two
different surfaces, say a hydrophobic plane and a hydrophilic
sphere, force ratio measurements displaying shear-dependant
results should be able to test whether the ideas put forward in
this paper are valid. Indeed, if the force decreases was really
due, not to bubbles, but to a change in the hydrodynamic boundary
condition for flow past the hydrophobic surface, the maximum force
decrease one could expect to obtain is $1/4$ for the case of a
perfectly slipping surface (see \cite{Vino95} for the calculation;
this result can also be found by symmetry about the plane where
slip occurs). If alternatively the measurements are due to a
``leaking mattress'' effect similar to the one we propose here,
equation \eqref{viscousforce} should also apply (with different
prefactors) and therefore so is equation \eqref{final};
consequently, force ratio smaller than $1/4$ should be obtained in
this case. This proposition for an experiment, together with the
prediction for the in-phase response of the force \eqref{elastic},
would allow our model to be tested experimentally.

\paragraph*{Acknowledgments}

We are grateful to Jacquie Ashmore, Jos\'e Gordillo, Steve
Granick, Jacob Israelachvili, Todd Squires, and Howard Stone for
useful discussions.

We also thank an anonymous referee for pointing out a mistake in
an earlier version of the manuscript. This research was supported
by the Harvard MRSEC, and the NSF Division of Mathematical
Sciences.

\newpage
\begin{table}[t]
\begin{tabular}{lcccc}
\hline\hline \\
& Ishida {\it et al.} \cite{nano1}\quad & \quad Tyrrell \& Attard \cite{nano2}\quad &\quad  Tyrrell \& Attard \cite{nano3}\quad &\quad Steitz {\it et al.} \cite{nano4} \\ \\
Projected area (nm$^2$) & $3.3\times 10^5$ & $4-6\times10^3$ &
$4-7 \times10^3$  & $2-11\times 10^3$ \\
Height\, $h_0$ (nm) & 40 & $20-30$ & $20-30$ & $<$ 18  \\
Radius of curvature\, $R_0$ (nm) & 1300 & $\sim 50$ & $40-60$ &
$30-100$ \\
Surface coverage $\phi$& $\sim 20\%$ & $\sim 100\%$ & $\sim 100\%$  &  $89\%$ \\
Macroscopic contact angle & 110\deg & 101\deg & 101\deg  &  $>$ 90\deg \\
Microscopic contact angle & 166\deg & $\sim120\,^\circ$ &  $117\,^\circ-130\,^\circ$ &  $130\,^\circ-147\,^\circ$\\  \\
 \hline\hline
\end{tabular}
\caption{Summary of experimental data on nanobubbles as found in
\cite{nano1,nano2,nano3,nano4} by atomic force microscopy:
projected area of each bubble on the solid surfaces, height above
the surfaces $h_0$, radius of curvature $R_0$, surface coverage
$\phi$, macroscopic and microscopic contact angle. The radius of
curvature and microscopic contact angles were inferred from the
other data assumming spherical cap nanobubbles.}\label{table}
\end{table}

\begin{thebibliography}{}
\bibitem{Goldstein}
{Goldstein S.}  {\it Modern Developments in Fluid Dynamics}, vol.
II  676 (Clarendon Press) (1938).

\bibitem{Granick2002}
{Zhu, X. and Granick, S.}  {\it Phys. Rev. Lett.} {\bf 88} 106102
(2002).

\bibitem{Jansons}
{Jansons, K.M.}  {\it Phys. Fluids} {\bf 31}  15 (1988).

\bibitem{Richardson}
{Richardson, S. }  {\it J. Fluid Mech.} {\bf 59}  707 (1973).

\bibitem{Pit}
{Pit, R., Hervert, H. and L\'eger, L.}  {\it Phys. Rev. Lett.}
{\bf 85} 980 (2000).

\bibitem{Baudry}
{Baudry, J., Charlaix, E., Tonck, A. and Mazuyer, D.}  {\it
Langmuir} {\bf 17}  5232 (2001).

\bibitem{Craig}
{Craig, V.S.J., Neto, C. and Williams, D.R.M.}  {\it Phys. Rev.
Lett.} {\bf 87}  054504 (2001).

\bibitem{Granick}
 {Zhu, X. and Granick, S.}  {\it Phys. Rev. Lett.} {\bf 87}
 096105 (2001).

\bibitem{Bonaccurso}
{Bonaccurso, E., Kappl, M. and Butt, H.-S.}  {\it Phys. Rev.
Lett.} {\bf 88}  076103 (2002).

\bibitem{Cheng}
{Cheng, J.-T. and Giordano, N.}  {\it Phys. Rev. E} {\bf 65}
 031206 (2002).

\bibitem{Meinhart}
{Tretheway, D.C. and Meinhart, C.D.}  {\it Phys. Fluids} {\bf 14}
 L9 (2002).

\bibitem{Cottin}
{Cottin-Bizonne, C., Jurine, S., Baudry, J., Crassous, J.,
Restagno, F. and Charlaix, E.}  {\it Eur. Phys. J. E} {\bf 9}
 47 (2002).

\bibitem{Nature}
{Thompson, P.A. and Troian, S.M.}  {\it Nature} {\bf 389} 360
(1997).

\bibitem{Barrat}
{Barrat, J.-L. and Bocquet, L.}  {\it Phys. Rev. Lett.} {\bf 82}
 4671 (1999).

\bibitem{BrennerPRE}
{Brenner, H. and Ganesan, V.}  {\it Phys. Rev. E} {\bf 61} 6879
(2000).

\bibitem{Koplik}
{Cieplak, M., Koplik, J. and Banavar, J.R.}  {\it Phys. Rev.
Lett.} {\bf 86}  803-806 (2001).

\bibitem{Denniston}
{Denniston, C. and Robbins, M.O.}  {\it Phys. Rev. Lett.} {\bf 87}
 178302 (2001).

\bibitem{Lauga}
{Lauga, E. and Stone, H.A.}  {\it J. Fluid Mech.}  {\bf 489} 55
(2003).

\bibitem{deGennes}
{de Gennes P.G.}  {\it Langmuir} {\bf 18}  3413 (2002).

\bibitem{Navier}
{Navier, C.L.M.H.}  {\it M\'emoires de l'Acad\'emie Royale des
Sciences de l'Institut de France} {\bf VI}  389 (1823).

\bibitem{nano1}
{Ishida, N., Inoue, T., Miyahara, M. and Higashitani, K.}
 {\it Langmuir} {\bf 16}  6377-6380 (2000).

\bibitem{nano2}
{Tyrrell, J.W.G. and Attard, P.}  {\it Phys. Rev. Lett.} {\bf 87}
 176104 (2001).

\bibitem{nano3}
{Tyrrell, J.W.G. and Attard, P.}  {\it Langmuir} {\bf 18} 160
(2002).

\bibitem{nano4}
{Steitz, R., Gutberlet T., Hauss, T., Kl\"osgen, B., Krastev, R.,
Schemmel, S., Simonsen, A.C. and Findenegg, G.H.} {\it Langmuir}
{\bf 19}  2409 (2003).

\bibitem{hydro0}
{Israelachvili J.}  {\it Intermolecular and Surface Forces}
(Academic Press) (1992).


\bibitem{hydro2}
{Carambassis, A., Jonker, L.C., Attard, P. and Rutland, M.W.}
 {\it Phys. Rev. Lett.} {\bf 80}  5357 (1998).

\bibitem{hydro3}
{Mahnke, J., Stearnes, J., Hayes, R.A., Fornasiero, D. and
Ralston, J.}  {\it Phys. Chem. Chem. Phys.} {\bf 1}, 2793 (1999).

\bibitem{hydro5}
{Attard, P.} {\it Langmuir} {\bf 16}   4455 (2000).

\bibitem{hydro6}
{Yakubov, G.E., Butt, H.-S. and Vinogradova, O.I.} {\it J. Phys.
Chem. B} {\bf 104}  3407 (2000).

\bibitem{Schnell}
{Schnell, E.}  {\it J. Appl. Phys.} {\bf 27}  1149 (1956).

\bibitem{Churaev}
{Churaev, N.V., Sobolev, V.D. and Somov, A.N.}  {\it J. Colloid.
Int. Sci.} {\bf 97}  574 (1984).

\bibitem{Watanabe}
{Watanabe, K., Udagawa, Y., and Udagawa, H.}  {\it J. Fluid Mech.}
{\bf 381}  225 (1999).

\bibitem{Vino95}
{Vinogradova, O.I.}  {\it Langmuir} {\bf 11}  2213 (1995).

\bibitem{lifetime}
{Ljunggren S. and Eriksson J.C.} {\it Colloids Surf. A} {\bf
129-130}  151 (1997).

\bibitem{attard}
{Attard P.} {\it Physica A} {\bf 314} (2002) 696.

\bibitem{Philip}
{Philip, J.R.} {\it J. App. Math. and Phys. (Zeitschrift f\"{u}r
Angewandte Mathematik und Physik)} {\bf 23}  353 (1972).

\bibitem{TaylorDispersion}
{Taylor, G.I.}  {\it Proc. Roy. Soc. A} {\bf  219}  186 (1953).

\bibitem{hocking}
{Hocking L.M.} {\it Phys. Fluids} {\bf 5} 793 (1994)

\bibitem{electrowetting}
Quilliet, C. \& Berge, B. {\it Curr. Opin. Colloid. Int. Sci.}{\bf
6} 34 (2001).

\bibitem{chou}
Chou T. {\it Phys. Rev. Lett.}, {\bf 87} 106101 (2001).

\bibitem{Onda}
{Onda, T., Shibuichi, S., Satoh, N. and Tsuji, K.}  {\it Langmuir}
{\bf 12}, 2125 (1996).

\bibitem{Bico}
Bico, J., Marzolin, C. and Qu\'er\'e, D.  {\it Europhys. Lett}
 {\bf 47}, 220 (1999).


\bibitem{Yakobov}
{Yakubov, G.E., Butt, H.-J. and Vinogradova, O.I.} {\it J. Phys.
Chem. B.} {\bf 104}  3407 (2000).

\end{thebibliography}
\end{document}